\documentclass[aoas,MSNbibl,nameyear,seceqn,dvips]{arximspdf}
\usepackage{dcolumn}
\usepackage{graphicx}

\doi{10.1214/11-AOAS530}
\volume{6}
\issue{4}
\pubyear{2012}
\firstpage{1499}
\lastpage{1530}

\makeatletter
\newcolumntype{d}[1]{D{.}{.}{#1}}
\makeatother

\begin{document}
\begin{frontmatter}

\title{A toolbox for fitting complex spatial point process models
using integrated nested Laplace approximation (INLA)}
\runtitle{Fitting complex spatial point process models with INLA}

\begin{aug}
\author[A]{\fnms{Janine B.} \snm{Illian}\corref{}\ead[label=e1]{janine@mcs.st-and.ac.uk}},
\author[B]{\fnms{Sigrunn H.} \snm{S{\o}rbye}}
\and
\author[C]{\fnms{H{\aa}vard} \snm{Rue}}
\runauthor{J. B. Illian, S. H. S{\o}rbye and H. Rue}
\affiliation{University of St Andrews, University
of Troms{\o} and Norwegian~University~of Science and Technology}
\address[A]{J. B. Illian\\
Centre for Research into Ecological\\ \quad and Environmental
Modelling\\
The Observatory\\ University of St Andrews\\
St Andrews KY16 9LZ\\ Scotland} %adresu isvedimo komanda gale!
\address[B]{S. H. S{\o}rbye\\
Department of Mathematics\\ \quad and Statistics\\ University of
Troms{\o}\\ 9037 Troms{\o}\\ Norway}
\address[C]{H. Rue\\
Department of
Mathematical Sciences\\ Norwegian University of Science\\ \quad and
Technology\\ 7491 Trondheim\\ Norway}
\end{aug}

% HISTORY:
\received{\smonth{10} \syear{2011}}
\revised{\smonth{11} \syear{2011}}

% ABSTRACT
%
\begin{abstract}
This paper develops methodology that provides a toolbox for
routinely fitting complex models to realistic spatial point
pattern data. We consider models that are based on log-Gaussian
Cox processes and include local interaction in these by
considering constructed covariates. This enables us to use
integrated nested Laplace approximation and to considerably speed
up the inferential task. In addition, methods for model comparison
and model assessment facilitate the modelling process. The
performance of the approach is assessed in a simulation study. To
demonstrate the versatility of the approach, models are fitted to
two rather different examples, a large rainforest data set with
covariates and a
point pattern with multiple marks.
\end{abstract}

% KEYWORDS
%

\end{frontmatter}

%s1 ###
\section{Introduction}\label{sec1}
%s1.1 ###
\subsection{Complex point process models}\label{sec1.1}
These days a large variety of complex statistical models can be fitted
routinely to complex data sets as a result of widely
accessible high-level statistical software, such as \texttt{R}
[\citet{R08}] or \texttt{winbugs} [\citet{winbugs}]. For
instance, the
nonspecialist user can estimate parameters in generalized linear
mixed models or run a Gibbs sampler to fit a model in a Bayesian
setting, and expert programming skills are no longer
required. Researchers from many different disciplines are now able to
analyze their data with sufficiently complex methods rather than
resorting to simpler yet nonappropriate methods. In addition, methods
for the assessment of a model's fit as well as
for the comparison of different models are widely used in practical
applications.

The routine fitting of spatial point process models to complex data
sets, however, is still in its infancy. This is despite a rapidly
improving technology that facilitates data collection, and a growing
awareness of the importance and relevance of small-scale spatial
information. Spatially explicit data sets have become increasingly
available in many areas of science, including plant ecology
[\citet{burslemal01}; \citet{lawal02}], animal ecology
[\citeauthor{forchal95}
(\citeyear{forchal95,forchal98})], geosciences [\citet{nayloral09}; \citet{ogata99}], molecular
genetics [\citet{hardyal02}], evolution [\citet
{johnsonal02}] and game
theory [\citet{killingbackal96}], with the aim of answering a similarly
broad range of scientific questions. Currently, these data sets are
often analyzed with methods that do not make full use of the available
spatially explicit information. Hence, there is a need for making
existing point process methodology available to applied scientists by
facilitating the fitting of suitable models.

In addition, real data sets are often more complex than the
classical data sets that have been analyzed with point process
methodology in the past. They often consist of the exact spatial
locations of the objects or events of interest, and of further
information on these objects, that is, potentially dependent
qualitative as well as quantitative marks or spatial covariates
[\citet{burslemal01}; \citet{mooreal10}]. There is an interest in fitting
complex joint models to the marks (or the covariates) as well as to
the point pattern. So far, the statistical literature has discussed
few examples of complex point process models of this type.

There have been previous advances in facilitating routine model
fitting for spatial point processes, in particular, for Gibbs processes.
Most markedly, the work by Baddeley and
Turner (\citeyear{badtur00}) has facilitated the routine fitting of Gibbs point
processes based on an approximation of the pseudolikelihood to avoid
the issue of intractable normalizing constants
[\citet{bermanturner92}; \citet{lawson92}] as well as the approximate
likelihood approach by \citet{huangal99}. Work by
\citet{baddeleyal05} and \citet{stoyanal91} has provided
methods for
model assessment for some Gibbs processes. Many of these have
been made readily available through the library \texttt{spatstat} for
\texttt{R} [\citet{baddeleyturner05}].

However, most Gibbs process models considered in the literature are
relatively simple in comparison to models that are commonly
used in the context of other types of data.
In an attempt to generalize the approach in \citet{baddeleyturner05},
\citet{illianal10} include random effects
in Gibbs point processes but more complex models, such as hierarchical
models or models including quantitative marks, currently cannot be
fitted in this framework. Similarly, methods for model comparison or
assessment considered in \citet{baddeleyal05} and \citet{stoyanal91}
are restricted to relatively simple models. Furthermore, both
estimation based on maximum likelihood and that based on pseudolikelihood
are approximate so that inference is not straightforward. The
approximations become less reliable with increasing interaction
strength [\citet{badtur00}].

Cox processes are another, flexible, class of spatial point process
models [\citet{moelleral07}], assuming a stochastic spatial trend
makes them particularly realistic and relevant in applications. Even
though many theoretical results have been discussed in the
literature for these
[\citet{moeller03}], the practical fitting of Cox point process models
to point pattern data remains difficult due to intractable
likelihoods. Fitting a Cox process to data is often based on Markov
chain Monte Carlo (MCMC) methods. These require expert programming
skills and can be very time-consuming both to tune and to run
[\citet{moeller03}] so that fitting complex models can easily become
computationally prohibitive. For simple models, fast minimum contrast
approaches to parameter estimation have been discussed
[\citet{moelleral07}].

However, approaches to routinely fitting Cox process models have
been discussed very little in the literature; similarly, methods for
model comparison or assessment for Cox processes have rarely been
discussed in the literature [\citet{illianrue10}; \citet{illianal11}]. To
the authors' knowledge, Cox processes have not been used outside the
statistical literature to answer concrete scientific questions.
Within the statistical literature Cox process models have focused on
the analysis of relatively small spatial patterns in terms of the
locations of individual species. Very few attempts have been made at
fitting models to both the pattern and the marks [\citet
{hoal08}; \citet{myllymaekial09}], in particular, not
to patterns with multiple dependent continuous marks, and joint models
of covariates and patterns have
not been considered.

This paper addresses two issues. It develops complex joint
models and, at the same time, provides methods facilitating the
routine fitting of these models. This provides a toolbox that allows
applied researchers to appropriately analyze realistic point pattern
data sets. We consider joint models of both the spatial pattern and
associated marks as well as of the spatial pattern and covariates.
Using a Bayesian approach, we
provide modern model fitting methodology for complex spatial point
pattern data similar to what is common in other areas of statistics
and has become a standard in many areas of application, including
methods for model comparison and validation. The approach is based
on integrated nested Laplace approximation (INLA) [\citet{rueal09}],
which speeds up parameter estimation substantially so that Cox
processes can be fitted within feasible time. In order to make the
methods accessible to nonspecialists, an \texttt{R} package that
may be used to run INLA is available and contains generic functions
for fitting spatial point process models; see
\url{http://www.r-inla.org/}.

%s1.2 ###
\subsection{Cox processes with local spatial structure}\label{sec1.2}
Applied researchers are aware that spatial behavior tends to vary at
a number of spatial scales as a result of different underlying
mechanisms that drive the pattern [\citet{wiegandal07};
\citet{latimeral09}].\vadjust{\goodbreak}
Local spatial behavior is often of specific
interest but the spatial structure also varies on a larger spatial
scale due to the influence of observed or unobserved spatial
covariates. Cox processes model spatial patterns relative to observed
or unobserved spatial trends and would be ideal models for these data
sets.

However, Cox processes typically do not consider spatial
structures at different spatial scales within the same model. More
specifically, a specific strength of
spatial point process models is their ability to take into account
detailed information at very small spatial scales contained in spatial
point pattern data, in terms of the local structure formed by an
individual and its neighbors. So far, Cox processes have often been
used to relate the locations of individuals to environmental
variation, phenomena that typically operate on larger spatial
scales. However, different mechanisms operate at a smaller spatial
scale. Spatial point data sets are often collected with a
specific interest in the local behavior of individuals, such as
spatial interaction or local clustering [\citet{lawal02}; \citet{latimeral09}].

We consider an approach to fitting Cox process models that reflects
both the local spatial structure and spatial behavior at a larger
spatial scale by using a constructed covariate together with spatial
effects that account for
spatial behavior at different spatial scales. This approach is
assessed in a simulation study and we also discuss issues specific
to this approach that arise when several spatial scales are
accounted for in a model.

This paper is structured as follows. The general methodology is
introduced in Section~\ref{Sec2}. In Section~\ref{constructcov} we
investigate the idea of mimicking local spatial behavior by using
constructed covariates in a simulation study in the context of
(artificial) data with known spatial structures and inspect patterns
resulting from the fitted models.
Section~\ref{rainforest} discusses a joint model of a large point
pattern and two empirical covariates along with a constructed covariate and
fits this to a rainforest data set. A hierarchical approach is
considered in Section~\ref{koalas}, where both (multiple) marks and
the underlying pattern are included in a joint model and fitted to a
data set of eucalyptus trees and koalas foraging on these trees.

%s2 ###
\section{Methods}\label{sec2}
\label{Sec2}

%s2.1 ###
\subsection{Spatial point process models}\label{sec2.1}

Spatial point processes have been discussed in detail in the
literature; see \citet{stoyal95}, \citet{lieshout00},
\citet{diggle03}, \citeauthor{moeller03} (\citeyear{moeller03,moelleral07}) and
\citet{book}. Here
we aim at modeling a spatial point pattern $\mathbf{x}=(\xi_1,
\ldots, \xi_n)$, regarding it as a realization from a spatial point
process $\mathbf{X}$. For simplicity we consider only point processes
in $\mathbb{R}^2$, but the approaches can be generalized to point
patterns in higher dimensions.

We refer the reader to the literature for information on different
(classes of) spatial point process models such as the simple Poisson
process, the standard null\vadjust{\goodbreak} model of complete spatial randomness, as
well as the rich class of Gibbs (or Markov) processes
[\citet{lieshout00}]. Here, we discuss the class of Cox
processes, in
particular, log-Gaussian Cox processes. Cox processes lend themselves
well to modeling spatial point pattern data with spatially varying
environmental conditions [\citet{moelleral07}], as they model spatial
patterns based on an underlying (or latent) random field
$\Lambda(\cdot)$ that describes the random intensity, assuming
independence given this field. In other words, given the random field,
the point pattern forms a Poisson process. Log-Gaussian Cox processes
as considered, for example, in \citet{moelleral98} and
\citeauthor{moeller03} (\citeyear{moeller03,moelleral07}), are a particularly flexible class,
where $\Lambda(s)$ has the form $\Lambda(s) = \exp\{
\mathfrak{Z}(s)\},$ and $\{\mathfrak{Z}(s)\}$ is a Gaussian random
field, $s \in\mathbb{R}^2$. Other examples of Cox processes include
shot-noise Cox processes [\citet{moeller03}].

Here, we consider a general class of complex spatial point process models
based on log-Gaussian Cox processes that allows the joint modeling of
spatial patterns along with marks and covariates. We include both small
and larger scale spatial behavior, using a constructed covariate and
additional spatial effects. The resulting models can be regarded as
latent Gaussian models and, hence, INLA can be used for parameter
estimation and model fitting.

%s2.2 ###
\subsection{Integrated nested Laplace approximation (INLA)}\label{sec2.2}
Cox processes are a special case of the very general class of
\textit{latent Gaussian models}, models of an outcome variable $y_i$
that assume independence conditional on some underlying latent field
$\boldsymbol{\zeta}$ and hyperparameters $\theta_j, j= 1, \ldots, J$.
\citet{rueal09} show that if $\boldsymbol{\zeta}$ has a sparse precision
matrix and the number of hyperparameters is small (i.e., $\leq$7),
inference based on INLA is fast.\looseness=-1

The main aim of the INLA approach is to approximate the posteriors of
interest, that is, the marginal posteriors for the latent field
$\pi(\zeta_i|\mathbf{y}),$ and the marginal posteriors for the
hyperparameters $\pi(\boldsymbol{\theta}_j|\mathbf{y})$, and use these to
calculate posterior means, variances, etc. These posteriors can be
written as
%
%e2.2 ###
%e2.1 ###
\begin{eqnarray}\label
{posterior1}
\pi(\zeta_i|\mathbf{y})& = &\int\pi(\zeta_i|\boldsymbol{\theta},
\mathbf{y})
\pi(\boldsymbol{\theta}|\mathbf{y})\,d\boldsymbol{\theta}, \\\label{posterior2}
\pi(\theta_j|\mathbf{y})&=& \int\pi(\boldsymbol{\theta}|\mathbf{y})\,d\boldsymbol{\theta}_{-j}.
\end{eqnarray}
The nested formulation is used to compute $\pi(\zeta_i|\mathbf{y})$ by
approximating $\pi(\zeta_i|\boldsymbol{\theta}, \mathbf{y})$ and
$\pi(\boldsymbol{\theta}|\mathbf{y})$, and then to use numerical integration
to integrate out $\boldsymbol{\theta}$. This is feasible, since the
dimension of $\boldsymbol{\theta}$ is small. Similarly,
$\pi(\theta_j|\mathbf{y})$ is calculated by approximating
$\pi(\boldsymbol{\theta}|\mathbf{y})$ and integrating out
$\boldsymbol{\theta}_{-j} $.

The marginal posterior in equations (\ref{posterior1}) and
(\ref{posterior2}) can be calculated using the Laplace approximation
\begin{displaymath}
\tilde{\pi}(\boldsymbol{\theta}|\mathbf{y}) \propto\frac{\pi
(\boldsymbol{\zeta},
\boldsymbol{\theta},
\mathbf{y})}{\tilde{\pi}_G(\boldsymbol{\zeta}|\boldsymbol{\theta
},\mathbf{y})
}  \bigg\vert_{ \boldsymbol{\zeta} =\boldsymbol{\zeta}^*(\boldsymbol{\theta})},\vadjust{\goodbreak}
\end{displaymath}
where $\tilde{\pi}_G(\boldsymbol{\zeta}|\boldsymbol{\theta},\mathbf{y})$ is
the Gaussian approximation to the full conditional of $\boldsymbol{\zeta}$,
and $\boldsymbol{\zeta}^*(\boldsymbol{\theta})$ is the mode of the full
conditional for $\boldsymbol{\zeta}$, for a given $\boldsymbol{\theta}$. This
makes sense, since the full conditional of a zero mean Gauss Markov
random field can often be well approximated by a Gaussian distribution
by matching the mode and the curvature at the mode
[\citet{rueheld05}]. Further details are given in \citet
{rueal09} who show that the nested
approach yields a very accurate approximation if applied to latent
Gaussian models. As a result, the time required for fitting
these models is substantially reduced.

%s2.3 ###
\subsection{Fitting log-Gaussian Cox processes with INLA}\label{sec2.3}
\label{lgc}

The class of latent Gaussian models comprises log-Gaussian Cox
processes and, hence, the INLA approach may be applied to fit
these. Specifically, the observation window is discretized into $N =
n_{\mathrm{row}} \times n_{\mathrm{col}}$ grid cells $\{ s_{ij}\}$, each with area
$|s_{ij}|$, $i = 1, \ldots, n_{\mathrm{row}}, j = 1, \ldots, n_{\mathrm{col}}$. The
points in the pattern can then be described by $\{\xi_{ijk_{ij}}\}$
with $k_{ij} = 1, \ldots, y_{ij}$, where $y_{ij}$ denotes the observed
number of points in grid cell $s_{ij}$. We condition on the point
pattern and, conditionally on $ \eta_{ij} = \mathfrak{Z}(s_{ij})$, we have
%
%e2.3 ###
\begin{equation}
\label{poisson} y_{ij}| \eta_{ij} \sim Po(|s_{ij}| \exp(\eta_{ij}));
\end{equation}
see \citet{rueal09}.

We model $\eta_{ij}$ as
%
%e2.4 ###
\begin{equation}
\eta_{ij} = \beta_0 + f(z_c(s_{ij})) + f_{1s}(s_{ij})+\cdots +f_{ps}(s_{ij}) + u_{ij}, \label{general}
\end{equation}
where the functions $f_{1s}(s_{ij})+\cdots +f_{ps}(s_{ij})$ are
spatially structured effects that reflect large scale spatial
variation in the pattern. These effects are modeled using a
second-order random walk on a lattice, using vague gamma priors for
the hyperparameter and constrained to sum to zero
[\citet{rueheld05}]. In the models that we discuss below, the
spatially structured effects relate to observed and unobserved
spatial covariates as discussed in the examples in Sections~\ref{rainforest} and~\ref{sec5}.
Including spatial covariates directly in the model as fixed effects
in addition to the random effects is straightforward. For
simplicity, we omit these in equation (\ref{general}) since this is
not relevant in the specific data sets and models discussed below.
$u_{ij}$ denotes a spatially unstructured zero-mean Gaussian i.i.d.\
error term, using a gamma prior for the precision.

Further, $z_c(s_{ij})$ denotes a constructed covariate. Constructed
covariates are summary characteristics defined for any location in
the observation window reflecting inter-individual spatial behavior
such as local interaction or competition. We assume that this
behavior operates at a smaller spatial scale than spatial
aggregation due to (observed or unobserved) spatial covariates, and
hence the spatially structured effects. The use of constructed
covariates yields models with local spatial interaction within the
flexible class of log-Gaussian Cox process models. It avoids issues
with intractable normalizing constants that are common in the
context of Gibbs processes [\citet{moeller03}], since the covariates
operate directly on the intensity of the pattern rather than on the
density or the conditional intensity [\citet{schoenberg05}].

The functional relationship between the outcome variable and the
constructed covariate is typically not obvious and might often not be
linear. We thus estimate this relationship explicitly by a smooth
function $f(z_c(s_{ij}))$ and inspect this estimate to gain further
information on the form of the spatial dependence. This function will
be modeled as a first-order random walk,
also constrained to sum to zero.

The constructed covariate considered in this paper is based on the nearest
point distance, which is simple and fast to compute. Specifically, for
each center point of the grid cells we
find the distance to the nearest point in the pattern outside this grid
cell as
%
%e2.5 ###
\begin{equation}\label{eq2.5}\label{nearestn}
z_c(s_{ij}) = d(s_{ij}) =\min_{\xi_l \in\mathbf{x}\setminus
{s_{ij}}}(\| c_{ij}-
\xi_l \|),
\end{equation}
where $c_{ij}$ denotes the center point of cell $s_{ij}$ and $\|
\cdot\|$ denotes the Euclidean distance. Defined this way, the
constructed covariate can be used both to model local repulsion and
local clustering.

During the modeling process, methods for model comparison based on
the deviance information criterion (DIC) [\citet
{spiegelhalter02}] may
be used to compare different models with different levels of
complexity. Furthermore, both the (estimated) spatially structured
field and the error field in
(\ref{general}) may be used to assess the model fit. The spatially
structured effect may be used
to reveal remaining spatial structure that is unexplained by the
current model and the unstructured effects may be
interpreted as a spatial residual. This provides a method for model
assessment akin to residuals in, for example,  linear models.

This approach yields a toolbox for fitting, comparing and assessing
realistically complex models of spatial point pattern data. We show
that different types of flexible models can be fitted to point
pattern data with complex structures using the INLA approach within
reasonable computation time. This includes joint models of large
point patterns and covariates operating on a large spatial scale and
local clustering (Section~\ref{rainforest}) as well as of a pattern
with several dependent marks which also depend on the pattern
(Section~\ref{koalas}).

%s2.4 ###
\subsection{Issues of spatial scale}\label{sec2.4}
In the natural world, different mechanisms operate at different
spatial scales [\citet{steffanal02}], and hence are reflected in
a spatial pattern
at these scales. It is crucial to bear this in mind during the
analysis of spatial data derived from nature, including spatial point
pattern data. Some mechanisms, such as seed dispersal in plants or
territorial behavior in animals, may operate at a local spatial
scale, while others, such as aggregation resulting from an
association with certain environmental covariates, operate on the
scale of the variation in these covariates, and hence often on a
larger spatial scale. In addition, a spatial scale that is relevant
in one application may not be relevant for a different data set.
Hence, the analysis of a spatial point pattern always involves a
consideration of the appropriate spatial scales at which mechanisms
of interest may operate, regardless of the concrete analysis methods.
Even as early as at the outset of a study, when an appropriately
sized observation window has to be chosen, relevant spatial scales
operating in the system of interest have to be taken into
consideration.

During the analysis the researcher has to carefully decide if
variation at a specific scale constitutes noise or whether it
reflects a true signal. It is hence crucial to be aware of which
mechanisms operate at which spatial scales prior to any spatial data
analysis. This may be done based on either background knowledge
(such as existing data on dispersal distances in plants or the sizes
of home ranges in territorial animals) or common sense.

In the models we discuss here, we explicitly take mechanisms
operating at several different scales into account and have to
choose these sensibly, based on knowledge of the systems. The
spatially structured effect reflects spatial autocorrelation at a
large spatial scale, whereas the constructed covariate is used to
describe small scale inter-individual behavior. In addition, since
we grid the data in this approach, the number of grid cells clearly
determines the spatial resolution, especially at a small scale, and
is clearly linked to computational costs and the extent to which
information is lost through gridding the data. In the following, we
discuss issues related to each of these three parts of the models
where spatial scale is relevant.

A spatially structured effect is typically included in a spatial
model as a spatially structured error term, that is, in order to
account for any spatial autocorrelation unexplained by covariates in
the model. INLA currently supports the 2nd order random walk on a
lattice as a model for this, with a gamma prior for the variance of
the spatially structured effect. The choice of this prior determines
the smoothness of the spatial effect and through this, the spatial
scale at which it operates. This prior has to be chosen carefully to
avoid overfitting. This is particularly crucial in the context of
spatial point patterns with relatively small numbers of points,
where the gridded data are typically rather sparse [\citet
{illianal11}]. If
the spatial effect is chosen to be too coarse, it explains the
spatial variation at too small a scale, resulting in a coarse
estimate of the spatially structured effect. This estimate would
perfectly explain every single data point, resulting in overfitting
rather than in a model of a generally interpretable trend. Given the
role of the spatially structured effect, it appears plausible to
choose the prior so that the spatial effect operates at a similar
spatial scale as the covariate. Problems can occur when the
spatially structured effect operates at a smaller scale than the
covariate, as it is then likely to explain the data better than the
covariates, rendering the model rather useless. In the absence of
covariate data, background knowledge on spatial scales may aid in
choosing the prior.

Small scale inter-individual spatial behavior is modeled by the
constructed covariate. As mentioned, this is done to account for
local spatial behavior if this is of specific interest in the
application. Again, there is a danger of overfitting, especially
since the constructed covariate is estimated directly from the data.
We discuss the practicality of using a spatial constructed covariate
in detail in Section~\ref{sec3} and only point out here that it has
to be carefully chosen, if possible with appropriate knowledge of
the specific system the data have been derived from.

The choice of prior for the spatially structured effect is strongly
related to the choice of grid size. However, in our experience the
overall results often do not change substantially when the grid size
was varied within reason. In applications, the locations of the
modeled objects as well as spatial covariates are sometimes given
on a grid with a fixed resolution. We recommend using a grid that is
not finer than that given by the data in the analysis.

%s3 ###
\section{Using a constructed covariate to account for local spatial
structure---a~simulation study}\label{sec3}
\label{constructcov}

In Section~\ref{rainforest} we use a constructed covariate primarily
to incorporate local spatial structure into a model, while
accounting for spatial variation at a larger spatial scale. To
illustrate the use of the given constructed covariate and to assess
the performance of the resulting models, we simulate point patterns
from various classical point-process models. Note, however, that we
do not aim at explicitly estimating the parameters of these models
but at assessing (i) whether known spatial structures may be
detected through the use of the constructed covariate, as suggested
here, and (ii) whether simulations from the fitted models generate
patterns with similar characteristics. In the applications we have
in mind, such as those discussed in the example in
Section~\ref{rainforest}, the data structure is typically more
complicated.

For the purpose of this simulation study we consider three different
situations: patterns with local repulsion (Section~\ref{repul}),
patterns with local clustering (Section~\ref{clust}) and patterns with
local clustering in the presence of a larger-scale spatial trend
(Section~\ref{clustinhom}). We generate example patterns from
different point process models with these properties on the unit
square. For all simulation results this observation window has been
discretized into a $100 \times100$ grid.

In Sections~\ref{repul} and~\ref{clust} we initially assume that there
is no large-scale spatial variation, with the aim of inspecting only
the constructed covariate, and we consider
%
%e3.1 ###
\begin{equation}
\eta_{ij}= \beta_0 + f(z_c(s_{ij}))\label{simplemodel},
\end{equation}
using the notation in Section~\ref{lgc}. In Section~\ref{clustinhom}
we consider both small- and large-scale spatial structures by including
a spatially structured effect $f_s(s_{ij})$ in addition to the
constructed covariate $z_c(s_{ij})$ and
%
%e3.2 ###
\begin{equation}
\eta_{ij}= \beta_0 + f(z_c(s_{ij})) + f_s(s_{ij}) \label{trendmodel}.
\end{equation}

To evaluate a fitted model, we apply the Metropolis algorithm
[\citet{metropolis53}] to simulate patterns from
these models and then compare characteristics of the simulated
patterns with the generated example patterns. More specifically, for
$i=1,\ldots, n_{\mathrm{row}}$ and $j=1,\ldots, n_{\mathrm{col}}$, denote the
joint distribution of $\mathbf{y}=\{y_{ij}\}$ given the latent field
$\boldsymbol{\eta}=\{\eta_{ij}\}$, by
\[
p(\mathbf{y}|\boldsymbol{\eta})=\prod_{i,j} p(y_{ij}|\eta
_{ij})=\prod_{i,j} \exp(-\lambda_{ij})\frac{\lambda
_{ij}^{y_{ij}}}{y_{ij}!},
\]
where the mean $\lambda_{ij}=|s_{ij}|\exp(\eta_{ij})$. For a given
example pattern, we first apply INLA to find the estimate $\hat
{\boldsymbol{\eta}}$ of the latent field for all grid cells. To evaluate
the estimated function of the constructed covariate for all arguments,
we apply the \texttt{splinefun} command in R to perform cubic spline
interpolation of the original data points. Using the Metropolis
algorithm, we assume an initial pattern $\mathbf{x}^{(0)}$, which is
randomly scattered in the unit square, having the same number of points
as the original pattern. The $k$th step of the algorithm is performed
by randomly selecting one point of the pattern $\mathbf{x}^{(k-1)}$
and proposing to move this point to a new position drawn uniformly in
the unit square. The proposal is accepted with probability
\[
\alpha=\min \biggl(1,\frac{p(\mathbf{y}^{(k)}|\hat{\boldsymbol
{\eta}})}{p(\mathbf{y}^{(k-1)}|\hat{\boldsymbol{\eta}})}
\biggr),\qquad k=1,2,\ldots,
\]
where $\mathbf{y}^{(k)}$ denotes the resulting grid cell counts for
$\mathbf{x}^{(k)}$. The simulated patterns in Sections~\ref{repul}--\ref{clustinhom} each result from $100\mbox{,}000$ iterations of the algorithm.

%f1 ###
\begin{figure}
\begin{tabular}{@{}c@{\hspace*{8pt}}c@{}}

\includegraphics{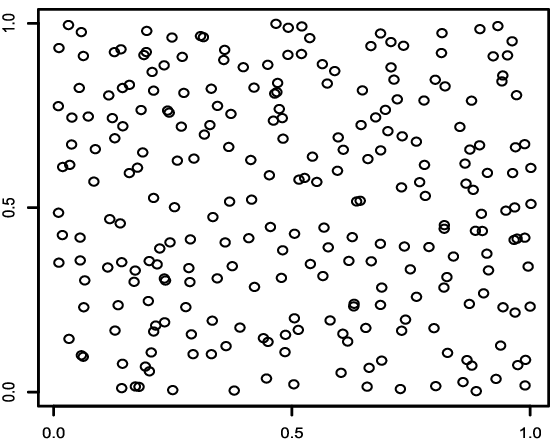}
&\includegraphics{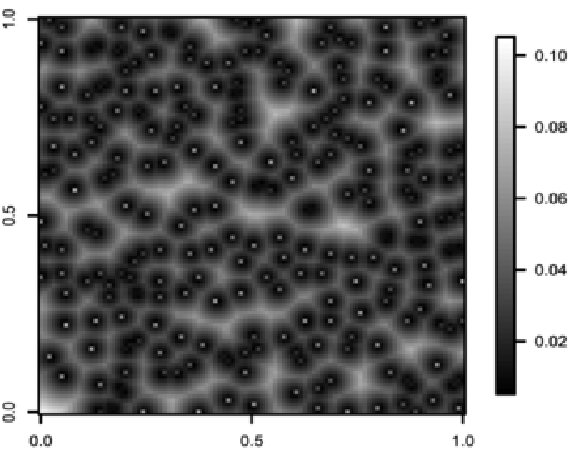}\\
(a)&(b)\\[6pt]

\includegraphics{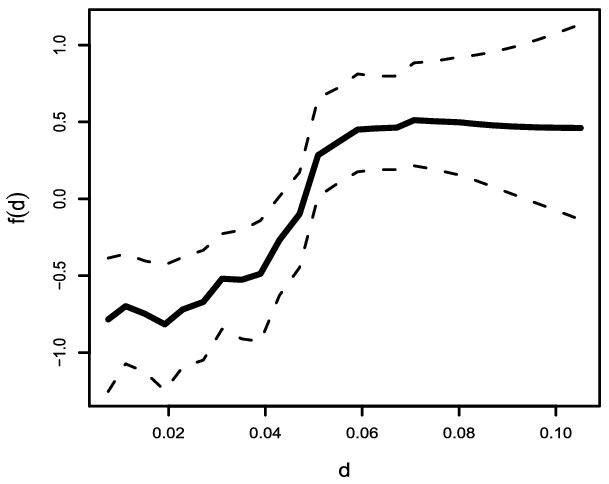}
&\includegraphics{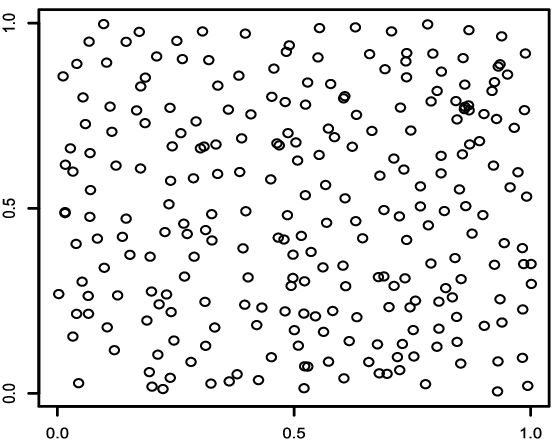}\\
(c)&(d)\\[6pt]

\includegraphics{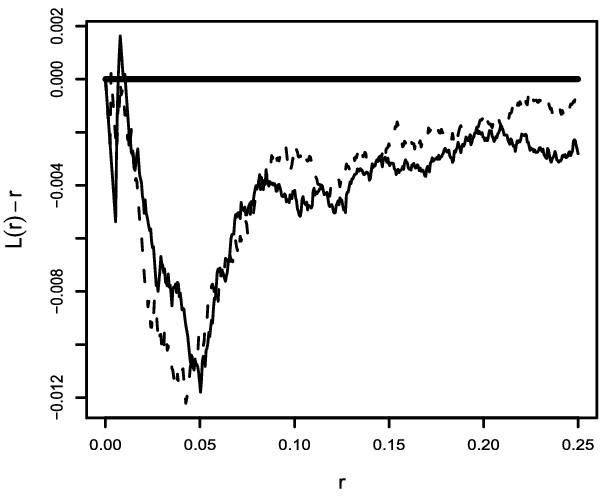}
&\includegraphics{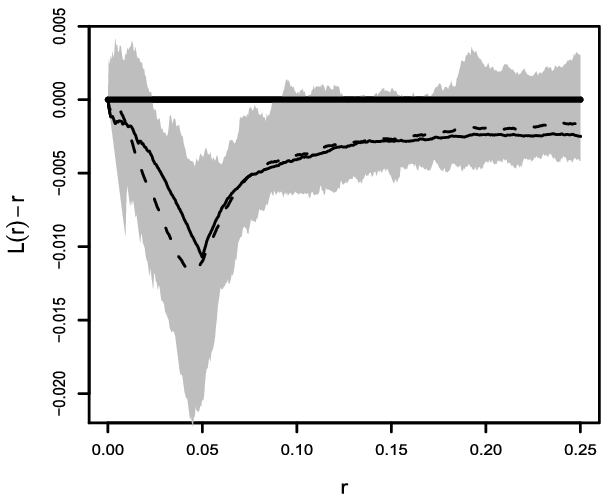}\\
(e)&(f)
\end{tabular}
\caption{Simulated Strauss process with medium repulsion
($r=0.05$, $\beta= 700$, $\gamma=0.05$) \textup{(a)}, the associated
constructed covariate for this pattern \textup{(b)}, the estimated
functional relationship between the outcome and the constructed
covariate \textup{(c)}, a pattern simulated from the fitted
model after 100,000 iterations \textup{(d)}, the estimated $L$-function for the
original pattern (solid line) and for the simulated pattern
(dashed line) \textup{(e)} and simulation envelopes for the $L$-function for 50
simulated patterns \textup{(f)}.}\label{straussplot}
\label{fig1}\end{figure}

%s3.1 ###
\subsection{Modeling repulsion}\label{sec3.1}
\label{repul}

To inspect the performance of the constructed covariate for repulsion,
we generate patterns from a homogeneous Strauss process
[\citet{strauss75}] on the unit square, with medium repulsion
$\beta
= 700$ (intensity parameter), $\gamma= 0.5$ (interaction parameter)
and interaction radius $r = 0.05$ [see Figure~\ref{straussplot}(a) for
an example]. We then fit a model to the pattern as in equation
(\ref{simplemodel}) using the constructed covariate in
(\ref{nearestn}) [Figure~\ref{straussplot}(b)]. The shape of the
estimated functional relationship between the constructed covariate
and the outcome variable is shown in Figure~\ref{straussplot}(c).
This function illustrates that the intensity in a grid cell is
influenced by the calculated distance in (\ref{nearestn}), as higher
distances will give higher intensities. Thus, the intensity is positively
related to the value of the constructed covariate, clearly reflecting repulsion.
At larger distances ($> $0.05) the function levels out distinctly,
indicating that beyond these distances the covariate and the
intensity are unrelated, that is, the spatial pattern shows random
behavior. In other words, the functional relationship not only
characterizes the pattern as regular but also correctly identifies
the interaction distance as 0.05.

The pattern resulting from the Metropolis--Hastings algorithm
[Figure~\ref{straussplot}(d)] shows very similar characteristics
to those in the original pattern. This indicates that the model
based on the nearest point constructed covariate in equation (\ref{eq2.5})
captures adequately the spatial information contained in the
repulsive pattern.

The estimated $L$-function [\citet{besag77Lfunction}] for the
simulated pattern and the original pattern confirm this impression,
as they look very similar [Figure~\ref{straussplot}(e)].
Additionally, we have calculated simulation envelopes for the
$L$-function of Strauss processes with the given parameter values,
using 50 simulated patterns and 100\mbox{,}000 iterations of the Metropolis algorithm
for each pattern [Figure~\ref{straussplot}(f)]. We notice that the
estimated $L$-functions of the original patterns are well within the
simulation envelopes for all distances.

%s3.2 ###
\subsection{Modeling clustering}\label{sec3.2}\label{clust}
\begin{figure}
\begin{tabular}{@{}c@{\hspace*{5pt}}c@{}}

\includegraphics{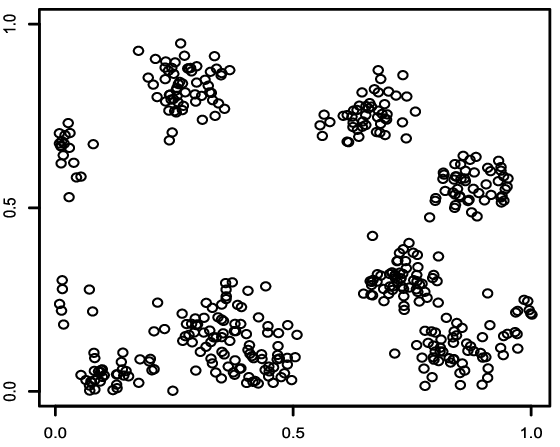}
&\includegraphics{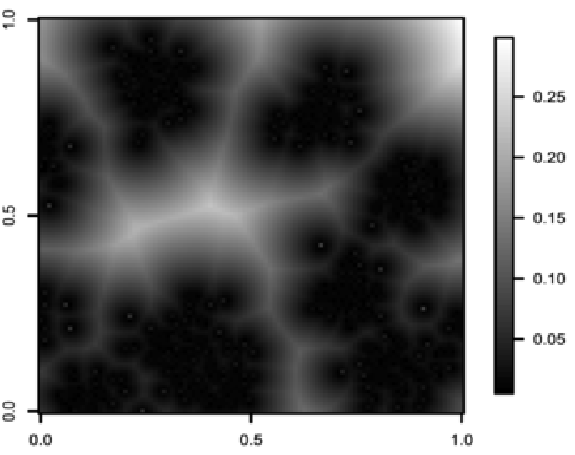}\\
(a)&(b)\\[6pt]

\includegraphics{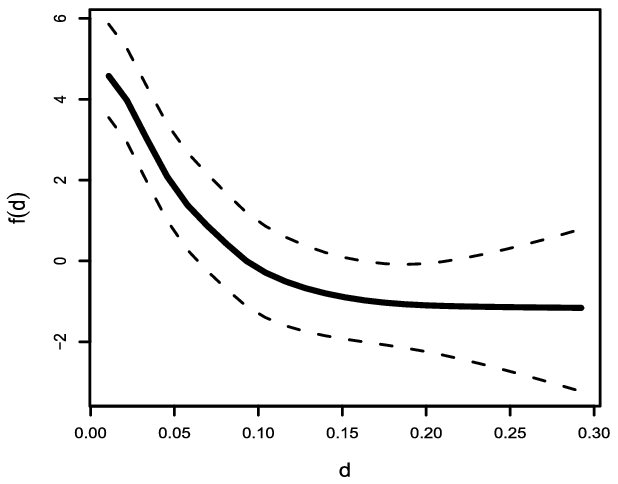}
&\includegraphics{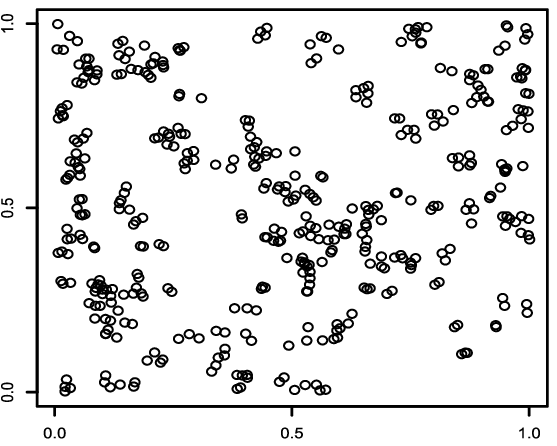}\\
(c)&(d)\\[6pt]

\includegraphics{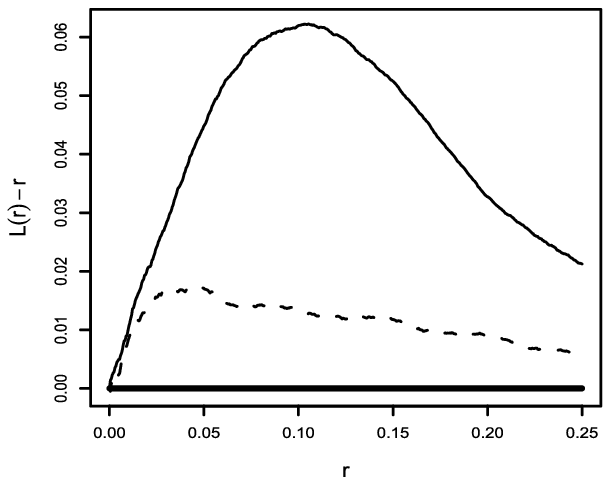}
&\includegraphics{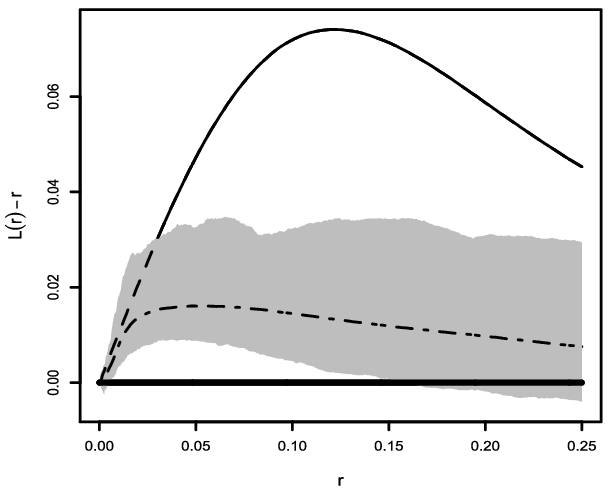}\\
(e)&(f)
\end{tabular}
\caption{Simulated Thomas process with parameters $\kappa= 10 $,
$\sigma= 0.05$ and $\mu=50$ \textup{(a)}, the associated constructed
covariate for this pattern \textup{(b)}, the estimated functional
relationship between the outcome and the constructed covariate
\textup{(c)}, a pattern simulated from the fitted model after 100\mbox{,}000
iterations \textup{(d)}, the estimated L-function for the original pattern
(solid line) and the simulated pattern
(dashed line) \textup{(e)} and simulation envelopes for the L-function for 50
simulated patterns \textup{(f)}.}\label{thomasplot}
\label{fig2}\end{figure}

In order to assess the performance of the model in
(\ref{simplemodel}) in the context of clustered patterns, we
generate patterns from a homogeneous Thomas process
[\citet{neymanal52}] in the unit square, with parameters $\kappa
= 10
$ (the intensity of the Poisson process of cluster centers), $\sigma
= 0.05$ (the standard deviation of the distance of a process point
from the cluster center) and $\mu= 50$ (the expected number of
points per cluster) [see Figure~\ref{thomasplot}(a) for an
example].
We fit the model in equation (\ref{simplemodel}) using the
constructed covariate in (\ref{nearestn}) [Figure~\ref{thomasplot}(b)]. The shape of the estimated functional relationship between the
constructed covariate and the outcome variable [Figure
\ref{thomasplot}(c)] now indicates that the intensity is
negatively related to the value of the constructed covariate as the
intensities increase for smaller distances, reflecting local
clustering. At larger
distances ($>$0.1) the function levels out, indicating that at these
distances the covariate and the intensity are unrelated.

%f2 ###

The pattern simulated from the fitted model [Figure
\ref{thomasplot}(d)] shows that the constructed covariate
introduces some clustering in the model. However, the resulting
pattern shows fewer and less distinct clusters than the original
pattern. Similarly, the estimated $L$-function
for the pattern simulated from the fitted model shows a weaker
local clustering effect than the original pattern [Figure
\ref{thomasplot}(e)]. This is also illustrated by the simulation
envelopes for 50 patterns of the fitted model which do not include
the true $L$-function [Figure~\ref{thomasplot}(f)].

%s3.3 ###
\subsection{Modeling small scale clustering in the presence of large-scale
inhomogeneity}\label{sec3.3}\label{clustinhom}

So far, we have considered constructed covariates only for patterns
with local interaction to illustrate their use. In
applications, however, different mechanisms operate at different
spatial scales. Patterns may be locally clustered, for example, due to
dispersal mechanisms, but may also show aggregation at a larger
spatial scale, for example, due to dependence on underlying observed or
unobserved covariates. Hence, the main reason for using constructed
covariates in the data example in Section~\ref{rainforest} is to
distinguish behavior at different spatial resolutions, in order to
provide information on mechanisms operating at different spatial
scales.

\begin{figure}
\begin{tabular}{@{}c@{\hspace*{8pt}}c@{}}

\includegraphics{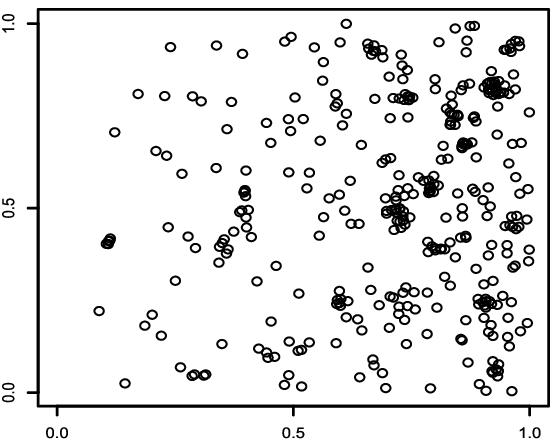}
&\includegraphics{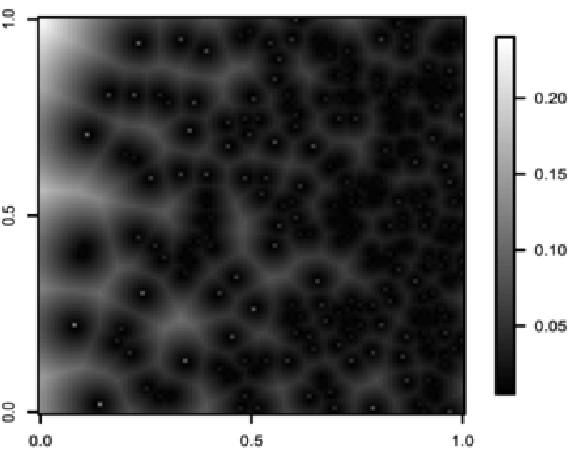}\\
(a)&(b)\\[6pt]

\includegraphics{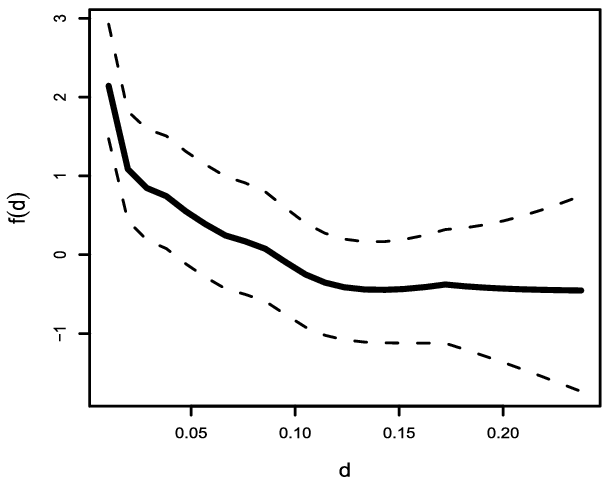}
&\includegraphics{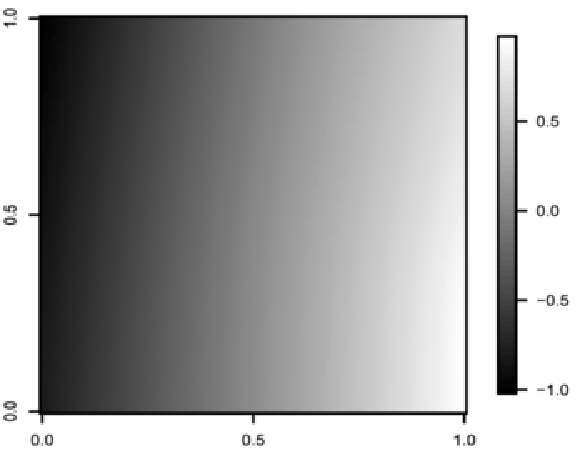}\\
(c)&(d)\\[6pt]

\includegraphics{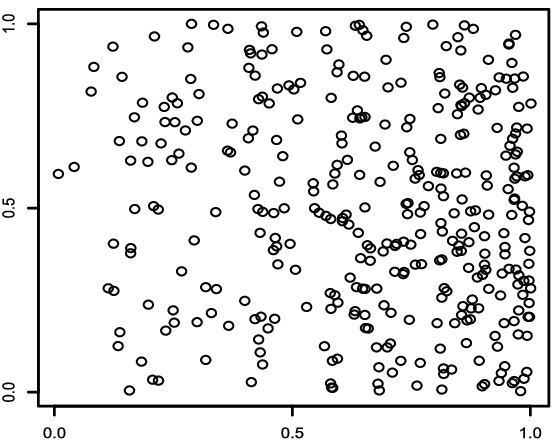}
&\includegraphics{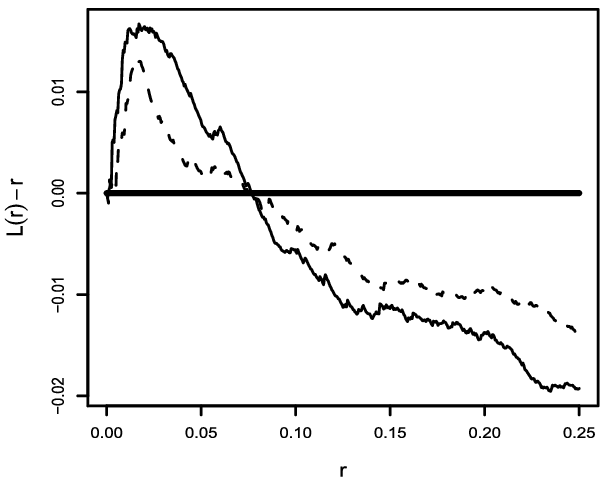}\\
(e)&(f)
\end{tabular}
\caption{Realization of an inhomogeneous Thomas process with
parameters $\sigma= 0.01$, $\mu= 5$ and trend function
$\kappa(x_1,x_2) = 50x_1$ superimposed on an inhomogeneous
Poisson process with trend function $\lambda={x_1}/4$ \textup{(a)}, the
associated constructed covariate for this pattern \textup{(b)}, the
estimated functional relationship between the outcome and the
constructed covariate \textup{(c)}, the estimated spatially
structured effect \textup{(d)}, a pattern simulated from the fitted
model after 100\mbox{,}000 iterations \textup{(e)} and the inhomogeneous L-function for
the original pattern (solid line) and the simulated pattern
(dashed line) \textup{(f)}.} \label{inhomThom}
\label{fig3}\end{figure}

We illustrate the use of constructed covariates in this context by
generating an inhomogeneous, locally clustered pattern mimicking a
situation where different mechanisms have caused local clustering
and large scale inhomogeneity. In applications, the inhomogeneity
may be modeled using suitable spatially varying covariates or
assuming an unobserved spatial variation or both. We generate patterns from
an inhomogeneous Thomas process with parameters $\sigma= 0.01$ and
$\mu= 5$ and a simple trend function for the intensity of parent
points given by $\kappa(x_1,x_2) = 50x_1$. Each pattern is then
superimposed with a pattern generated from an inhomogeneous Poisson
process with trend function $\lambda=x_1/4$ [Figure~\ref{inhomThom}(a)].

%f3 ###

We again use the constructed covariate in (\ref{nearestn}) [see Figure
\ref{inhomThom}(b)] and fit the model in (\ref{trendmodel}). The
inspection of the functional relationship between the constructed
covariate and the outcome [Figure~\ref{inhomThom}(c)] shows that at
small values of the covariate the intensity is negatively related to
the constructed covariate, reflecting clustering at smaller
distances. The estimated
spatially structured effect picks up the larger-scale spatial
behavior [Figure~\ref{inhomThom}(d)]. Patterns simulated from the
fitted model look quite similar to
the original pattern [Figure~\ref{inhomThom}(e)].
However, local clustering is slightly stronger
in the original pattern than in the simulated pattern [Figure
\ref{inhomThom}(f)].

\begin{figure}

\includegraphics{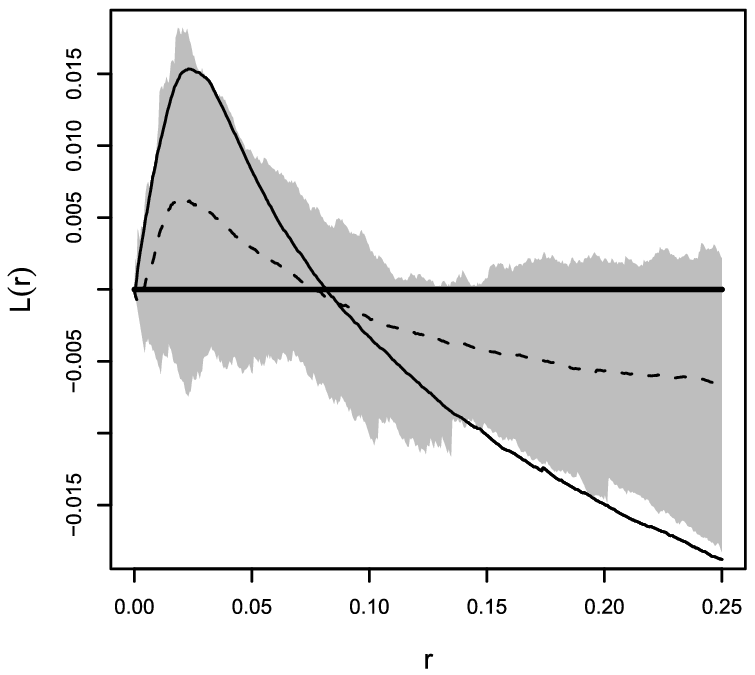}

\caption{Inhomogeneous Thomas process. Simulation envelopes for 50
patterns generated from the fitted model using 100\mbox{,}000 iterations, the
inhomogeneous
$L$-function for a Poisson process (bold solid line), the mean of the
inhomogeneous $L$-function for the generated (solid) and simulated
(dashed) patterns.} \label{inhomThomse}\vspace*{-3pt}
\label{fig4}\end{figure}

This is again confirmed by the simulation envelopes for the simulated
patterns from the fitted model, as shown in Figure~\ref{inhomThomse}.
The mean estimated L-function for the generated patterns is very close
to the upper edge of the simulation envelopes and partly outside,
indicating that the fitted model does not reflect the strength of
clustering sufficiently well.
%
%f4 ###

%s3.4 ###
\subsection{Discussion on constructed covariates}\label{sec3.4}
With the aim of assessing the performance of models with constructed
covariates reflecting small scale inter-individual spatial behavior,
we consider a number of simulated point patterns for three different
scenarios: repulsion, clustering and small-scale clustering in the
presence of large scale inhomogeneity. In all cases, the local spatial
structure can be clearly identified. The constructed covariate
does not only take account of local spatial structures but also
characterizes the spatial behavior. The functional form of the
dependence of the intensity on the constructed covariate clearly
reflects the character of the local behavior.\vadjust{\goodbreak}

This section presents only a small part of an extensive simulation
study; the results shown here are typical examples. We have run
simulations from the same models as above with different sets of
parameters and have obtained essentially the same results. Further,
fitting the model in equation (\ref{simplemodel}) to patterns
simulated from a homogeneous Poisson process resulted in a
nonsignificant functional relationship, that is, the modeling approach
does not pick up spurious clustering or regularity.

The approach allows us to fit models that take into account small-scale
spatial behavior, regularity as well as clustering, in the
context of log-Gaussian Cox processes, that is, as latent Gaussian
models. Since these can be fitted using the INLA approach, fitting is
fast and exact. In addition, we avoid some of the typical problems
that arise with Gibbs process models, that is, we do not face issues of
intractable normalizing constants, and regular as well as clustered
patterns may be modeled.

However, the simulations also show that the approach of using
constructed covariates works clearly better with repulsive patterns
than with clustered patterns. This is akin to similar issues with Gibbs
processes, where repulsive patterns are less problematic to model than
clustered patterns. Certainly, this is related to the fact that it is
difficult to tell apart clustering from inhomogeneity [\citet
{diggle03}]. When working with constructed covariates the issues
highlighted, that is, that local clustering may have been
underestimated, have to be taken into account, especially in the
interpretation of results.\vadjust{\goodbreak}

Certainly, the constructed covariate in equation (\ref{nearestn})
that we consider here is not the only possible choice. A covariate
based on distance to the nearest point is likely to be rather
``short-sighted,'' so that other constructed covariates might be more
suitable for detecting specific spatial structures. In particular,
taking into account these limitations, it is not surprising that
patterns simulated from models show less clustering
than the original data. More general covariates such as
the distance to the $k$th nearest point may be considered. Other
covariates, such as the local intensity or the number of points
within a fixed interaction radius from a location $s \in\mathbb
R^2$, are certainly also suitable. A nice property of the given
constructed covariate based on nearest-point distance is that it is
parameter-free. For this reason, it is not necessary to choose
explicitly the resolution of the local spatial behavior, for example, as
an interaction radius. Also, note that since the distance to the
nearest point in point pattern $\mathbf{x}$ for a location $s \in
\mathbb R$ may be interpreted as a graph associated with $\mathbf{x}
\cup\{s\}$, other constructed covariates based on different types
of graphs [\citet{rajalaal11}] may also be used as constructed
covariates. Similarly, an approach based on morphological functions
may be used for this purpose. Note that one could also consider
constructed marks based on first or second order summary
characteristics [\citet{book}] that are defined only for the
points in
the pattern and include these in the model.

Distinguishing spatial behavior at different spatial scales is
clearly an ill-posed problem, since the behavior at one spatial scale
is not independent of that at different spatial scales
[\citet{diggle03}]. The approach we take here will not always be able
to distinguish clustering at different scales. However, different
mechanisms that operate at very similar spatial scales are likely to
be nonidentifiable by any method, irrespective of the choice of model
or the constructed covariate. Constructed covariates hence
only provide useful results when the processes they are meant to
describe operate
at a spatial scale that is distinctly smaller than the larger scale
processes in the same model.

Admittedly, the use of constructed covariates is of a rather
subjective and  ad hoc nature. Clearly, in applications the
covariates have to be constructed carefully, depending on the
questions of interest; different types of constructed covariates may
be suitable in different contexts. However, similarly subjective
decisions are usually made when a model is fitted that is purely
based on empirical covariates, as these have been specifically
chosen as potentially influencing the outcome variable, based on
background knowledge. In addition, due to the apparent danger of
overfitting, constructed covariates should only be used if there is
an interest in the local spatial behavior in a specific data set
and if there is reason to believe that small- and large-scale
spatial behavior are operating at scales that are different enough
to make them identifiable.

%s4 ###
\section{Joint model of a point pattern and environmental
covariates}\label{sec4}\label{rainforest}

%s4.1 ###
\subsection{Modeling approach}\label{sec4.1}

In this example we consider a point pattern $\mathbf{x} = (\xi_1,
\ldots, \xi_n)$, where the number of points $n$ is potentially very
large and several spatial covariates have been measured. The point
pattern is assumed to depend on one or several (observed or
unobserved) environmental covariates for which data $z_{1}, \ldots,
z_{p}$ exist. In the application that we have in mind the values of
these have been observed in a few locations that are typically
different from locations of the objects that form the pattern. In
previous modeling attempts the values of the covariates in the
locations of the objects are then either interpolated or modeled
separately so that (estimated) values are used for locations were
the covariates have not been observed. However, these covariates are
likely to have been collected with both sampling and measurement
error. In the specific case we consider here (see Section
\ref{rainforestex}) they concern soil properties, which are measured
much less reliably than the topography covariates in models such as
those in \citet{waagepetersen07}, \citet{waagepetersenal09}. In addition, it is
less clear for soil variables than for topography covariates if these
influence the presence of trees, or whether the presence of trees
impacts on the soil variables. Whereas models in which the soil
variables are considered fixed and not modeled alongside the pattern,
the model we deal with here does not make any assumption on the
direction of this influence.

As a result, we suggest a joint model of the covariates along with
the pattern that uses the original (noninterpolated) data on the
covariates and accounts for measurement error. That is, we fit the
model in equation (\ref{general}) to $\mathbf{x}$ and jointly fit a
model to the covariates. The pattern and the covariates are linked
by joint spatial fields. An additional spatially structured effect
is used to detect any remaining spatial structures in the pattern
that cannot be explained by the joint fields with the covariates.

In the case of $p = 2$ we fit the following
model, where the pattern is modeled as
%
%e4.1 ###
\begin{equation}
\eta_{ij} = \beta_{0} + f(z_{c}(s_{ij})) + f_s(s_{ij}) + g_s(s_{ij})
+ h_s(s_{ij}),
\label{raineq1}
\end{equation}
and the covariates as
%
%e4.2 ###
\begin{equation}
z_{1ij} = f_s(s_{ij}) + u_{ij},
\label{raineq2}
\end{equation}
and
%
%e4.3 ###
\begin{equation}
z_{2ij} = g_s(s_{ij}) + v_{ij},
\label{raineq3}
\end{equation}
where $z_{1ij}$ and $z_{2ij}$ are the observed covariates in grid
cells where the covariates have been measured and missing where they
have not been measured. $f(z_{c}(s_{ij}))$ represents the function of
the constructed covariate
(\ref{nearestn}). $f_s(\cdot)$ and $g_s(\cdot)$
are spatially structured effects, that is, reflect a random field for
each of the covariates and $h_s(\cdot)$ reflects spatial autocorrelation
in the pattern unexplained by the covariates; $u_{ij}$
and $v_{ij}$ are spatially unstructured fields used to account
for measurement or sampling error.

In addition to the spatial effect reflecting the empirical covariates,
which are likely to have an
impact on the larger scale spatial behavior, we use the constructed
covariate to account for local clustering. In the application we
have in mind (see Section~\ref{rainforestex}) this clustering is a
result of seed-dispersal mechanisms operating on a much smaller spatial
scale than that of the aggregation of individuals due to an association
with environmental covariates.

%s4.2 ###
\subsection{Application to example data set}\label{sec4.2}\label{rainforestex}
%s4.2.1 ###
\subsubsection{The rainforest data}\label{sec4.2.1}

Some extraordinarily detailed multi-species maps are being collected
in tropical forests as part of an international effort to gain greater
understanding of these ecosystems [\citet
{condit98}; \citet{hubbellal99}; \citet{burslemal01}; \citet{hubbellal05}]. These data comprise
the locations of
all trees with diameters at breast height (dbh) 1 cm or greater, a
measure of the size of the trees (dbh), and the species identity of
the trees. The data usually amount to several hundred thousand trees
in large (25 ha or 50 ha) plots that have not been subject to any
sustained disturbance such as logging. The spatial distribution of
these trees is likely to be determined by both spatially varying
environmental conditions and local dispersal.

Recently, spatial point process methodology has been applied to
analyze some of these data sets [\citet{lawal09}; \citet{wiegandal07}] using
nonparametric descriptive methods as well as explicit models
[\citet{waagepetersen07}; \citet{guan08}; \citet{waagepetersenal09}; \citet{yueal11}]. \citet
{rueal09} model
the spatial pattern formed by a tropical rain forest tree species on
the underlying environmental conditions and use the INLA approach to
fit the model.

\begin{figure}
\begin{tabular}{@{}c@{\hspace*{8pt}}c@{}}

\includegraphics{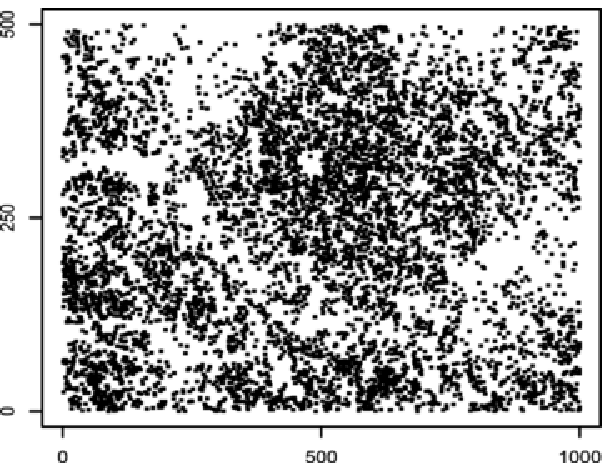}
&\includegraphics{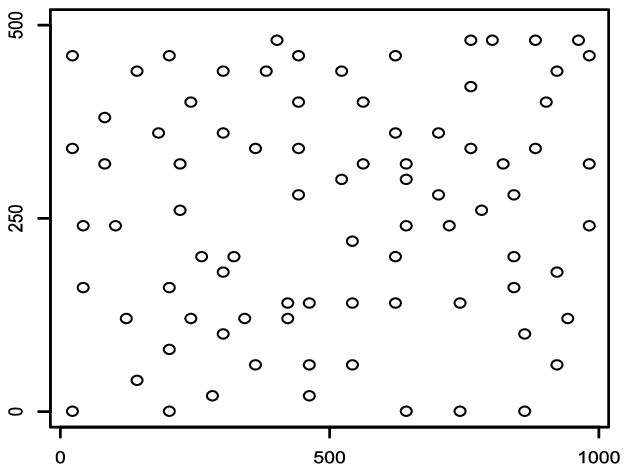}\\
(a)&(b)
\end{tabular}
\caption{Spatial pattern of the species \emph{Aporusa microstachya}
in Pasoh Forest Reserve, Peninsular Malaysia and locations where soil
variables have been measured.} \label{patternrain}
\label{fig5}\end{figure}

We analyze a data set that is similar to those discussed in the above
references. Since the spatial structure in a forest reflects dispersal
mechanisms as well as association with environmental conditions, we
include a constructed covariate to account for local clustering. The
model is fitted to a data set from a 50 ha
forest dynamics plot at Pasoh Forest Reserve, Peninsular Malaysia.
This study focuses on the species \textit{Aporusa microstachya}
consisting of 7416 individuals [Figure~\ref{patternrain}(a)]. The
environmental covariates have been observed in 83 locations that are
distinct from the locations of the trees [Figure~\ref{patternrain}(b)]. The plot lies in a forest that has never been logged with very
narrow streams on almost flat land. The data collected in 1995 are
used here when the plot contained 320\mbox{,}903 stems from 817 species. The
species is the most common small tree on the plot. It is of interest
if this species, as an aluminium accumulator, covaries with
magnesium availability, as aluminium uptake might constrain its
capacity to take up nutrient cations such as magnesium. In addition,
its covariation with phosphorus is considered here as the element
thought to be the nutrient primarily limiting forest productivity
and individual tree growth in tropical forests [Burslem, personal
communication (February 2011)].

%f5 ###

%f6 ###
\begin{figure}
\begin{tabular}{@{}c@{\hspace*{13pt}}c@{}}

\includegraphics{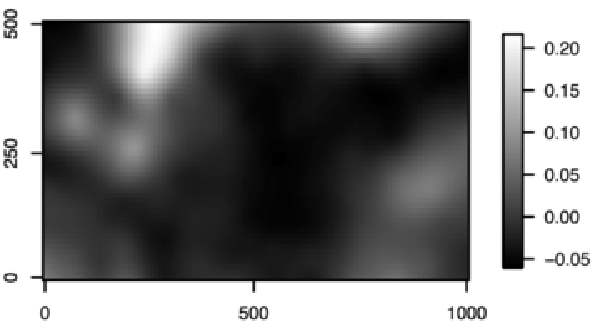}
&\includegraphics{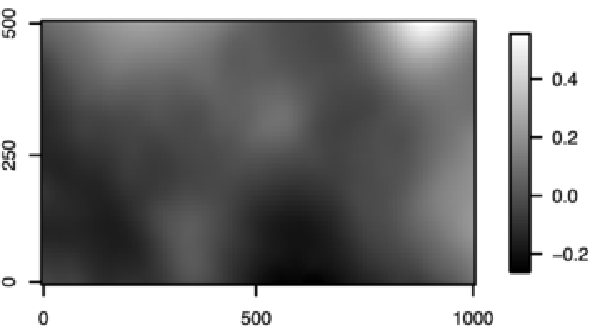}\\
(a)&(b)\\[6pt]

\includegraphics{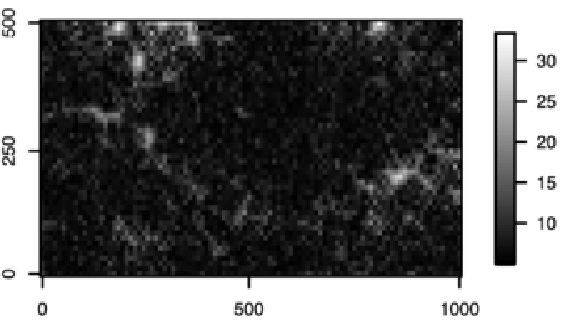}
&\includegraphics{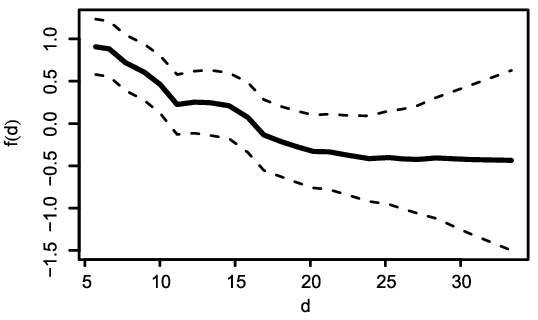}\\
(c)&(d)\\[6pt]

\includegraphics{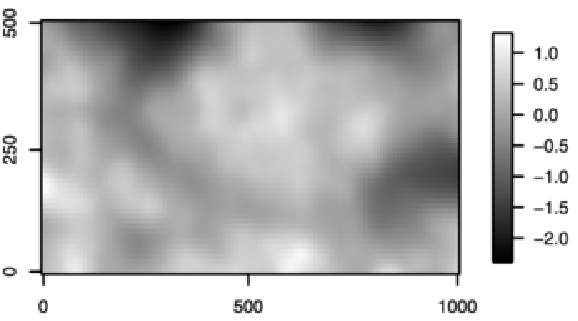}
&\includegraphics{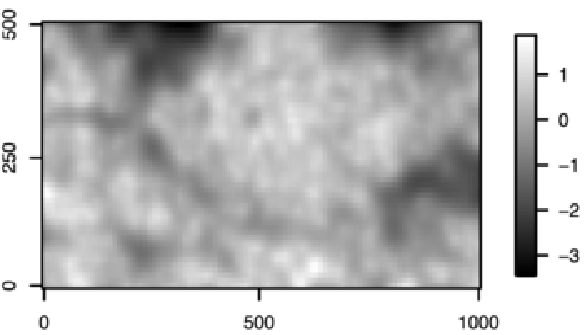}\\
(e)&(f)
\end{tabular}
\caption{Rainforest data. Top panels: The estimated spatially
structured effect for the covariates phosphorus \textup{(a)} and
magnesium \textup{(b)}. Middle panels: The calculated constructed covariate \textup{(c)}
and the
estimated function of the constructed covariate \textup{(d)}. Bottom panels:
The estimated spatially structured effect for the pattern with \textup{(e)} and
without the constructed covariate term in the model \textup{(f)}.}
\label{covrain}
\label{fig6}\end{figure}

%s4.2.2 ###
\subsubsection{Results}\label{sec4.2.2}
We run the full model as described in equations (\ref{raineq1}) to
(\ref{raineq3}), in which the observation area is discretized into
$50\times100$ grid cells. The spatial effect of the two empirical
covariates, phosphorus $f_s(\cdot)$ and magnesium $g_s(\cdot)$, are displayed
in Figure~\ref{covrain}(a) and (b). We notice that these effects are
very smooth, but we have to remember that the covariate information is
sparse and only available in $83$ grid cells. In terms of DIC, the
empirical covariate terms explain some spatial structure of the pattern
as DIC increases from 15\mbox{,}379 to 15\mbox{,}440 if these two terms are not
included. High phosphorus seems to coincide with low tree density and a
similar, but less clear, pattern emerges for magnesium. Currently, the
ecological literature cannot explain these results, but they could be
related to resource partitioning along axes of soil nutrient
availability [Burslem, personal communication (September 2011),
\citet{johnal07}]. In addition, it is currently also unclear if
the soil properties cause an aggregation of trees, as they provide
suitable growing conditions, or whether a high tree intensity leads to
low levels of magnesium or phosphorus resulting from the chemical
composition of the leaf litter.

The plot of the constructed covariate in Figure~\ref{covrain}(c)
illustrates the resolution of the local clustering represented by it.
The resulting estimated function of the constructed covariate is shown
in Figure~\ref{covrain}(d), which indicates that it accounts for
clustering of up to a distance of 15 metres.
The estimated spatial\vadjust{\goodbreak} effect $h_s(\cdot)$ for the pattern is given in
Figure~\ref{covrain}(e), while Figure~\ref{covrain}(f) displays the
estimated spatially structured effect if the constructed covariate is
left out of the full model. This last figure shows clear local
structure in the spatial effect and might give a model which is
overfitted to the actual pattern. Including the constructed covariate,
the local structure of the spatial effect is removed, making the
spatial effect smoother. This indicates that spatial behavior at a
local scale has been picked up by the constructed covariate. In this
way the model can account for spatial structures at different scales.
The two unstructured spatial fields in equations (\ref{raineq2}) and
(\ref{raineq3}) do not show any particular pattern (results not
shown). Fitting this model took 55 minutes to run (2.66 GHz Intel Core
i7 processor).

%s4.3 ###
\subsection{Discussion on rainforest data}\label{sec4.3}
In this section we consider a log Gaussian Cox process model and
fit it jointly to a point pattern data set with a large number of
points and two covariates that have been observed at a relatively
small number of points within the plot. \citet{waagepetersen07} and
\citet{waagepetersenal09} model the patterns formed by rainforest
tree species with this data structure, using Thomas processes to
include local clustering resulting from seed dispersal. This
approximate approach is based on the minimum contrast method for
parameter estimation. \citet{rueal09} consider the same data in the
context of Cox processes to demonstrate that log-Gaussian Cox
processes can be fitted conveniently to a large spatial point
pattern using INLA relative to environmental covariates which are
assumed to be known everywhere and fixed. In many typical applications,
however, the values of spatial covariates in the location of the
points forming the point process are not known. Similarly, the
direction of the relationship between soil properties and tree presence
may be not clear. We generalize the
approach in \citet{rueal09} here and fit a joint model of the
pattern and the covariates. This approach distinguishes between
locations where the values of the covariates are available but
potentially subject to measurement error and those where they are not.
In addition, it does not assume that the soil variables impact on the
pattern but not vice versa.
We also consider a constructed covariate that reflects local clustering
as a
result of local seed dispersal, as discussed above.

The given approach accommodates model comparison and model assessment, both
of which are of practical value in many applications. An inspection of
the estimated spatially structured effect in Figure
\ref{covrain}(e) indicates that some spatial structure still
remains in the point pattern which cannot be explained by the
current model, that is, the current model can still be improved on.
Hence, judging by Figure~\ref{covrain}(e), it might be possible
to improve the model by including further covariates and the
structure of the estimated spatial effect might be used to suggest a
suitable covariate. Previous approaches to fitting a model to these
data [\citet{waagepetersen07}; \citet{waagepetersenal09}] neither have been able
to reveal the shortcomings
of the models nor to provide mechanisms that help identify covariates
that might improve the model.

The function of a constructed covariate [Figure~\ref{covrain}(d)],
which reflects local clustering up to a distance of 15 metres, may be
interpreted as a seed dispersal kernel. Biological research has shown
that this species is likely to be dispersed primarily by small
understorey birds that feed in the canopy and mostly drop the seeds
beneath the parent tree. Since trees of the species \textit{Aporusa
microstachyathese} are relatively small, 15 m reflect the maximum
radius of the tree crown [\citet{burslemal01}].

The approach discussed here can be extended easily to allow more
complex models to be fitted, such as a model of both the spatial
pattern and
associated marks, along the lines of the model discussed in Section
\ref{koalas}. For instance, this may include a model of both the
spatial pattern and the size and the growth of the trees. Here, both
size and growth might depend
on the spatial pattern and growth might also depend on size.

%s5 ###
\section{Modeling marks and pattern in a marked point pattern with
multiple marks}\label{sec5}
\label{koalas}

Modeling the behavior of individuals in space based simply on the
individuals' locations and ignoring their properties is certainly a
gross over-simplification for many systems. In practice, researchers
hence often collect data on the locations of the individuals along
with data on additional properties, that is, marks. In this section we
discuss a marked point pattern with several dependent marks, which
also depend on the spatial pattern, and consider a joint model of the
marks and the pattern. Models where marks depend on the point pattern
have recently been considered in the literature [\citet
{menezes05}; \citet{hoal08}; \citet{myllymaekial09}]. Also note the work by
\citet{diggleal10}, where a point process with intensity dependent
marks is used in the context of preferential sampling in
geostatistics. The model we fit here is more general than these
related models, since we model multiple dependent marks jointly with
the pattern.

%s5.1 ###
\subsection{Data structure and modeling approach}\label{sec5.1}
We analyze a spatial point pattern $\mathbf{x}= (\xi_1, \ldots,
\xi_n)$ together with several types of nonindependent associated
marks. We consider only two marks $\mathbf{m}_1 = (m_{11}, \ldots,
m_{1n})$ and $\mathbf{m}_2= (m_{21}, \ldots, m_{2n})$ here, but the
approach can be generalized in a straightforward way to include more
than two marks. The $\mathbf{m}_1$ are assumed to follow an
exponential family distribution $F_{1\theta_1}$ with parameter vector
$\boldsymbol{\theta}_{1}= (\theta_{11},\ldots, \theta_{1q})$ and to depend
on the intensity of the point pattern, while the $\mathbf{m}_2$ are
assumed to follow a (different) exponential family distribution
$F_{2\theta_{2}}$ with parameter vector $\boldsymbol{\theta}_{2}=
(\theta_{21},\ldots, \theta_{2q})$ and to depend both on the
intensity of
the point pattern and on the marks $\mathbf{m}_1$. Without loss
of generality, the parameters $\theta_{11}$ and $\theta_{21}$ are the
location parameters of the distributions $F_1$ and $F_2$,
respectively.

We discretize the observation window as discussed in Section
\ref{lgc}, and for the spatial pattern we assume the model
%
%e5.1 ###
\begin{equation}
\eta_{ij} = \beta_{01} + f(z_c(s_{ij}))+\beta_1 \cdot f_s(s_{ij}) +
u_{ij},\label{koalaeqn1}
\end{equation}
using the same notation as in (\ref{general}). For the marks, we
construct a model where the marks $\mathbf{m}_1$ depend on the pattern
by assuming that they depend on the same spatially structured effect
$f_s(s_{ij})$. Specifically, we assume that $m_{1}(\xi_{ijk_{ij}})|
\kappa_{ijk_{ij}} \sim
F_{1\theta_1}(\kappa_{ijk_{ij}},\theta_{12},\ldots, \theta_{1q} )$
with
%
%e5.2 ###
\begin{equation}
\kappa_{ijk_{ij}} = \beta_{02} + \beta_2 \cdot f_s(s_{ij}) + v_{ijk_{ij}},
\label{koalaeqn2}
\end{equation}
where $v_{ijk_{ij}}$ is another error term. The marks $\mathbf{m}_2$
are assumed to depend both on the spatial pattern through
$f_s(s_{ij})$ and on the marks $\mathbf{m}_1$. We thus have that
$m_{2}(\xi_{ijk_{ij}})| \nu_{ijk_{ij}} \sim
F_{2\theta_2}(\nu_{ijk_{ij}}, \theta_{22},\ldots, \theta_{2q})$ with
%
%e5.3 ###
\begin{equation}
\nu_{ijk_{ij}} = \beta_{03} + \beta_3 \cdot f_s(s_{ij}) + \beta_4
\cdot
m_1(\xi_{ijk_{ij}}) + w_{ijk_{ij}}, \label{koalaeqn3}
\end{equation}
where $w_{ijk_{ij}}$ denotes another error term.

%f7 ###
\begin{figure}[b]
\begin{tabular}{c}

\includegraphics{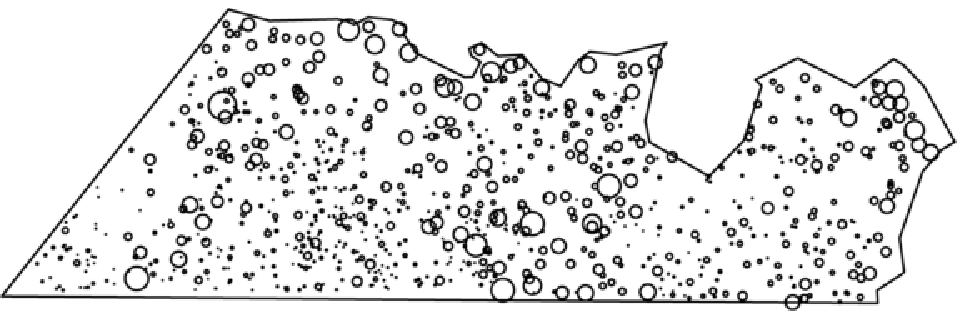}
\\
(a)\\[6pt]

\includegraphics{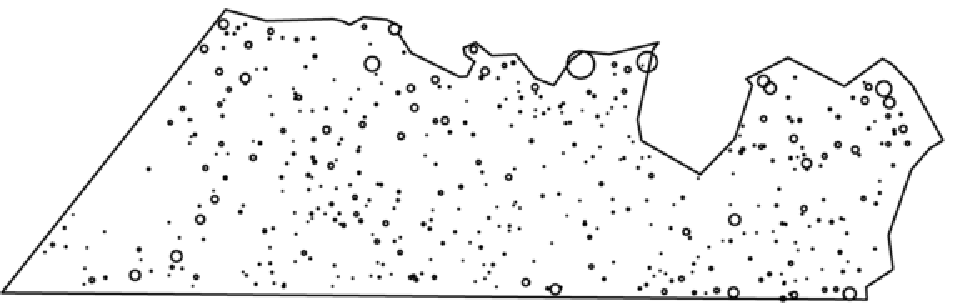}
\\
(b)
\end{tabular}
\caption{Spatial pattern formed by the locations of the eucalyptus
trees in the koala data set; the diameters of the circles
reflect the value of the leaf marks \textup{(a)} and the frequency
marks \textup{(b)}, respectively.} \label{koalaspnew}
\label{fig7}\end{figure}

%s5.2 ###
\subsection{Application to example data set}\label{sec5.2}

%s5.2.1 ###
\subsubsection{Koala data}\label{sec5.2.1}

Koalas are arboreal marsupial herbivores native to Australia with a
very low metabolic rate. They rest motionless for about 18 to 20 hours
a day, sleeping most of that time. They feed selectively and live
almost entirely on eucalyptus leaves. Whereas these leaves are
poisonous to most other species, the koala gut has adapted to digest
them. It is likely that the animals preferentially forage leaves that
are high in nutrients and low in toxins as an extreme example of
evolutionary adaptation. An understanding of the koala-eucalyptus
interaction is crucial for conservation efforts [\citet{mooreal10}].

The data have been collected in a study conducted at the Koala
Conservation Centre on Phillip Island, near Melbourne, Australia.
For each of 915 trees within a reserve enclosed by a koala-proof
fence (Figure~\ref{koalaspnew}), information on the leaf chemistry
and on the frequency of koala visits has been collected.\vadjust{\goodbreak} The leaf
chemistry is summarized in a measure of the palatability of the
leaves (``leaf mark'' $\mathbf{m}_L$). Palatability is assumed to
depend on the intensity of the point pattern. In addition,
``frequency marks'' $\mathbf{m}_F$ describe for each tree the diurnal
tree use by individual koalas collected at monthly intervals between
1993 and March 2004. The $\mathbf{m}_F$ are assumed to depend on the
intensity of the point pattern as well as on the leaf marks.

There are no additional covariate data available for the given data
set. Hence, for the locations of the trees we use the model in
(\ref{koalaeqn1}) with notation as above. For the leaf and
frequency marks we use the models in equations
(\ref{koalaeqn2}) and (\ref{koalaeqn3}), respectively. The leaf
marks are assumed to follow a normal distribution and the frequency
marks a Poisson distribution, that is, $m_{L}(\xi_{ijk_{ij}})|
\kappa_{ijk_{ij}} \sim N (\kappa_{ijk_{ij}},\sigma^2)$ and
$m_{F}(\xi_{ijk_{ij}})| \nu_{ijk_{ij}} \sim
Po(\exp(\nu_{ijk_{ij}}))$.

%s5.2.2 ###
\subsubsection{Results}\label{sec5.2.2}
With these distributional assumptions for the marks, we fit a joint
model as given in equations (\ref{koalaeqn1})--(\ref{koalaeqn3}) to
the data set. The results are based on an observation window
discretized into 1571 grid cells. In order to fit spatial effects, we
embed this area within a rectangular area. For the constructed
covariate, we perform a simple edge correction for the distances in
(\ref{nearestn}), assuming missing values in grid cells in which the
distance from the center point to the border is shorter than the
nearest-point distance.

%t1 ###
\begin{table}[b]
\tabcolsep=0pt
\caption{DIC values and computation time for different fitted models
for the koala data}
\label{tabledickoala}
\label{tab1}\begin{tabular*}{290pt}{@{\extracolsep{\fill}}lld{5.0}d{3.0}@{}}
\hline
\textbf{Model} & \multicolumn{1}{l}{\textbf{Terms}} & \multicolumn{1}{c}{\textbf{DIC}} & \multicolumn{1}{c@{}}{\textbf{Time (s)}}\\
\hline
1. & Only error terms & 11\mbox{,}308& 4\\
2. & Add intercepts & 8362 & 4\\
3. & Add fixed covariate ($\beta_4$) & 7640& 5\\[3pt]
4. & Add spatial effect & & \\
&  -- Only for pattern & 7511& 25 \\
&  -- For pattern and leaf marks & 7312 & 71 \\
&  -- For pattern and frequency marks & 7193 & 61 \\
&  -- For pattern and both marks (final model) & 6943 & 142 \\[3pt]
5. &Add constructed covariate & 6943 & 189 \\
\hline
\end{tabular*}
\end{table}

When fitting complex models it can be useful to apply a stepwise
procedure to study the impact of each term in the model. Table~\ref{tab1}
illustrates DIC-values and computation time (in seconds) of models of
increasing complexity. In the first three steps we initially run a
model with only error terms and then add intercepts and the fixed
covariate for the frequency marks. Step 4 illustrates the effect of
adding the spatial effect $f_s(\cdot)$ in modeling the pattern together
with one or both of the two marks, in which DIC decreases to 6943.
Inclusion of the constructed covariate in (\ref{koalaeqn1}) does not
improve the model fitting for this data set. This is not surprising, as
the original pattern does not seem\vadjust{\goodbreak} to exhibit any strong local
clustering effect and as a result the estimated function of the
constructed covariate is not significantly different from 0.

%f8 ###
\begin{figure}
\begin{tabular}{cc}

\includegraphics{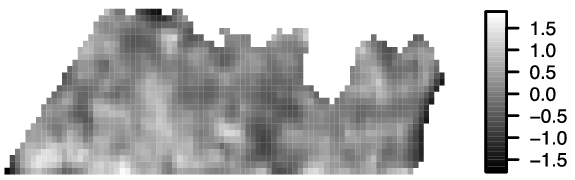}
&\includegraphics{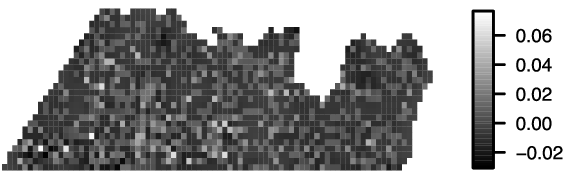}\\
(a)&(b)\\[6pt]

\includegraphics{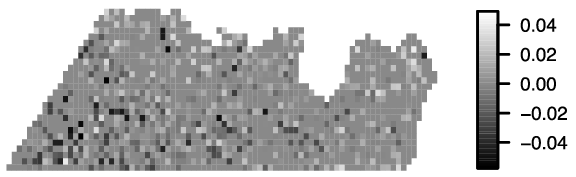}
&\includegraphics{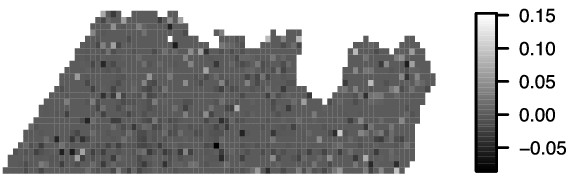}\\
(c)&(d)
\end{tabular}
\caption{Plots of the estimated common spatial effect \textup{(a)} and the three
unstructured effects $u_{ij}$, $v_{ijk_{ij}}$ and
$w_{ijk_{ij}}$ \textup{(b)}--\textup{(d)} for the koala data
set.} \label{koalaresnew}
\label{fig8}\end{figure}

%t2 ###
\begin{table}[b]
\tablewidth=180pt
\caption{Posterior means and $95\%$ credible intervals for parameters
in the koala model}
\label{tableparam}
\label{tab2}\begin{tabular}{@{}lcc@{}}
\hline
\textbf{Parameter} & \textbf{Mean} & $\boldsymbol{95\%}$ \textbf{credible interval} \\
\hline
$\beta_2$ & $-$1.18& $[-1.39,-0.96]$ \\
$\beta_3$ & \hphantom{$-$}1.72 & $[1.45, 1.98]$\\
$\beta_4$ &\hphantom{$-$}1.38 &$[1.24, 1.52]$ \\
\hline
\end{tabular}
\end{table}

The estimated common spatial effect [Figure~\ref{koalaresnew}(a)]
represents spatial autocorrelation present in the pattern and the marks
which might be the result of related environmental processes such as
nutrient levels in the soil. The estimated parameter value for $\beta
_2$ and $\beta_3$ have opposite signs (Table~\ref{tab2}). The negative sign for
$\beta_2$ indicates that palatability is low where the trees are
aggregated, which might have been caused by competition for soil
nutrients in these areas. The positive sign for $\beta_3$ reflects
that the koalas are more likely to be present in areas with higher
intensity. Recalling that the data have been accumulated over time,
this might be due to the koalas being more likely to change from one
tree to a neighboring tree where the trees are aggregated. The mean of
the posterior density for the parameter $\beta_4$ in the final model
is 1.38, indicating a significant positive influence of palatability on
the frequency of koala visits to the trees. The three unstructured
terms are given in Figures~\ref{koalaresnew}(b)--(d).
A slight trend in the residuals for the leaf marks may be observed in
Figure~\ref{koalaresnew}(c), with lower values toward the bottom left
probably reflecting an inhomogeneity that cannot be accounted for by
the joint spatial effect $f_s(s_{ij})$.

%s5.3 ###
\subsection{Discussion of koala data}\label{sec5.3}
The example considered in this section is a marked Cox process
model, that is, a model of both the spatial pattern and two types of
dependent marks, providing information on the spatial pattern at
the same time as on the marks and their dependence. In cases
where the marks are of primary scientific interest, one could view
this approach as a model of the marks which implicitly takes the
spatial dependence into account by modeling it alongside the marks.
The model we use here is similar to approaches taken in
\citet{menezes05}, \citet{hoal08}, \citet{myllymaekial09}. Since our approach is
very flexible, it can easily be generalized to allow for separate
spatially structured effects for the pattern and the marks and to
include additional empirical covariates; these have not been
available here. Hence, using the approach considered here, we are
able to fit easily a complex spatial point process model to a marked
point pattern and to
assess its suitability for a specific data set.

Marked point pattern data sets where data on marks are likely to
depend on an underlying spatial pattern are not uncommon. Within
ecology, for instance, metapopulation data [\citet{hanskial97}]
typically consist of the locations of subpopulations and their
properties, and have a similar structure to the data set considered
here. These data sets may be modeled using a similar approach and it
is straightforward to fit related but more complex models, including
empirical covariates or temporal replicates. Similarly, marks are
available for the rainforest data discussed in Section~\ref{sec4}. As mentioned
there, a model that includes the marks of the trees
may also be fitted using the approach discussed here.

%s6 ###
\section{Discussion}\label{sec6}
Researchers outside the statistical community have become familiar
with fitting a large range of different models to complex data sets
using software available in \texttt{R}. This paper provides a very
flexible framework
for routinely fitting models to complex spatial point pattern data
with little computational effort using models that account for both
local and global spatial
behavior.
We consider complex data examples and demonstrate how marks as well as
covariates can be included in a joint model. That is, we consider a
situation where the marks and the covariates can be modeled along
with the pattern and show that it is computationally feasible to do so.
We can take account of local spatial structure by using a
constructed covariate, which we discuss in detail in Section
\ref{constructcov}.

The two models discussed here indicate that our approach can be
applied in a wide range of situations and is flexible enough to
facilitate the fitting of other even more complex models. It is
feasible to fit several related
models to realistically complex data sets if necessary, and to use
the DIC to aid the choice of covariates. The posterior distributions
of the estimated parameters can be used to assess the significance of
the influence of different covariates in the models. Through the use
of a structured spatial effect and an unstructured spatial effect it
is possible to assess the quality of the model fit. Specifically, the
structured spatial effect can be
used to reveal spatial correlations in the data that have not been
explained with the covariates and may help researchers identify
suitable covariates to incorporate into the model. Spatially
unstructured effects may be used to account for and identify extreme
observations such as locations where covariate values have been
collected with a particularly strong measurement error.

There is an extensive literature on descriptive and nonparametric
approaches to the analysis of spatial point patterns, specifically
on (functional) summary characteristics describing first and second
order spatial behavior, in particular, on Ripley's $K$-function
[\citet{ripley76}] and the pair correlation function
[\citet{stoyal95}]. In both the statistical and the applied
literature these have been discussed far more frequently than
likelihood based modeling approaches, and provide an elegant means
for characterizing the properties of spatial patterns [Illian et
al. (\citeyear{book})]. A~thorough analysis of a spatial point pattern typically
includes an extensive exploratory analysis and in many cases it may
even seem unnecessary to continue the analysis and fit a spatial
point process model to a pattern. An exploratory analysis based on
functional summary characteristics, such as Ripley's $K$-function or
the pair-correlation function, considers spatial behavior at a
multitude of spatial scales, making this approach particularly
appealing. However, with increasing complexity of the data, it
becomes less obvious how suitable summary characteristics should be
defined for these, and a point process model may be a suitable
alternative. For example, it is not obvious how one would jointly
analyze the two different marks together with the pattern in the
koala data set based on summary characteristics. However, as
discussed in Section~\ref{koalas}, it is straightforward to do this
with a hierarchical model. In addition, most exploratory analysis
tools assume the process to be first-order stationary or at least
second-order reweighted stationary [Baddeley et al. (\citeyear{B00})]---a
situation that is both rare and difficult to assess in applications,
in particular, in the context of realistic and complex data sets. The
approach discussed here does not make any assumptions about
stationarity but explicitly includes spatial trends into the model.

In the literature, local spatial behavior has often been modelled
by a Gibbs process. Large-scale spatial behavior may be
incorporated into a Gibbs process model as a parametric or
nonparametric, yet deterministic, trend, while it is treated as a
stochastic process in itself here. Modeling the spatial trend in a
Gibbs process hence often assumes that an explicit and deterministic
model of the trend as a function of location (and spatial
covariates) is known [\citet{baddeleyturner05}]. Even in the
nonparametric situation, the estimated values of the underlying
spatial trend are considered fixed values, which are subject
neither to stochastic variation nor to measurement error. Since it
is based on a latent random field, the approach discussed here
differs substantially from the Gibbs process approach and assumes a
hierarchical, doubly stochastic structure. This very flexible class
of point processes provides models of local spatial behavior
relative to an underlying large-scale spatial trend. In realistic
applications this spatial trend is not known. Values of the
covariates that are continuous in space are typically not known
everywhere and have been interpolated. It is likely that spatial
trends exist in the data that cannot be accounted for by the
covariates. The spatial trend is hence not regarded as
deterministic but assumed to be a random field. This approach allows
to jointly model the covariate and the spatial pattern as in the
model used for the rainforest example data set. Clearly, unlike
Gibbs processes, log Gaussian Cox processes do not allow second order
inter-individual interactions to be included in a model. In a
situation where these are of primary interest, Cox processes are
certainly not suitable.

In order to make model fitting feasible, the continuous Gaussian
random field is approximated here by a discrete Gauss Markov random
field. While this is computationally elegant, one might argue that
this approximation is not justified and is too coarse, resulting in
an unnecessary loss of information. Clearly, since any model only
has a finite representation in a computer, model fitting approaches
often work with some degree of discretization. However, and more
importantly, \citet{lindgrenal11} show that there is an explicit
link between a large class of covariance functions (and hence the
Gaussian random field based on these) and Gauss Markov random fields,
clearly pointing out that the approximation is indeed justified. In
addition, based on the results discussed in \citet{lindgrenal11}, the
approaches taken in this paper may be extended to avoid the
computationally wasteful need of having to use a regular grid
[\citet{illiansimpson11}]. \citet{illianal11} also mention
the issue
of complex boundaries structures that are particularly relevant for
point process data sets where the observation window has been chosen
to align with natural boundaries that may impact on pattern. While
this is clearly not an issue for the rainforest data set since the
boundaries have been chosen arbitrarily, the koala data set,
however, has been observed in an observation window surrounded by a
koala proof fence. This fence does probably not impact on the locations of
the trees nor the leaf chemistry but might increase the frequency of
koala visits near the fence. The approach in \citet{lindgrenal11}
may be used to define varying boundary conditions for different
parts of the data set, and hence allow for more realistic modeling
for data sets with complicated boundary structures.

In summary, the methodology discussed here, together with the
\texttt{R} library \texttt{R-INLA}
(\url{http://www.r-inla.org/}), makes complex spatial point
process models accessible to scientists outside the statistical
sciences and provides them with a toolbox for routinely fitting and
assessing the fit of suitable and realistic point process models to
complex spatial point pattern data.

\section*{Acknowledgments}
Some of the ideas relevant to the rainforest data were developed during
a working group on ``Spatial analysis of tropical forest biodiversity''
funded by the Natural Environment Research Council and English Nature
through the NERC Centre for Population Biology and UK Population
Biology Network. The data were collected by the Center for Tropical
Forest Science and the Forest Institute of Malaysia funded by the U.S.
National Science Foundation, the Smithsonian Tropical Research
Institution and the National Institute of Environmental Studies
(Japan).

We would like to thank David Burslem, University of Aberdeen, and
Richard Law, University of York, for introducing the rainforest data
into the statistical community and for many in-depth discussions over
the last few years. We also thank Colin Beale, University
of York, and Ben Moore, James Hutton Institute, Aberdeen, for
extended discussions on the koala data.

The authors also gratefully acknowledge the financial support of
Research Councils UK for Illian.

%suskaldyti doi

% imsref loaded by smiklovaite, 2012-03-29 11:06:55
% imsref loaded by smiklovaite, 2012-03-29 11:27:46
%

\printaddresses

\end{document}